\documentstyle[bo99,epsfig,psfig]{article}

\title{BEPPOSAX OBSERVATIONS OF MARKARIAN 501 IN JUNE 1999 }
\author{E.~Pian$^1$,
L.~Chiappetti$^2$,
P.~Giommi$^4$,
F.~Tavecchio$^3$,
L.~Maraschi$^3$,
E.~Palazzi$^1$,
F.~Aharonian$^5$,
M.~Catanese$^6$, 
A.~Celotti$^7$,
B.~Degrange$^8$,
A.~Djannati-Atai$^8$,
G.~Fossati$^9$,
G.~Ghisellini$^3$,
H.~Krawczynski$^5$,
C.~M.~Raiteri$^{10}$,
R.~M.~Sambruna$^{11}$, 
D.~Smith$^{12}$,
G.~Tagliaferri$^3$,
G.~Tosti$^{13}$,
A.~Treves$^{14}$, 
C.~M.~Urry$^{15}$,
M.~Villata$^{10}$}
\affil{ $^1$ ITESRE/CNR, Bologna, Italy,
$^2$ IFCTR/CNR, Milan, Italy,
$^3$ Astronomical Obs. of Brera, Milan, Italy,
$^4$ SAX Science Data Center, Rome, Italy,
$^5$ Max-Planck-Institut f\"ur Kernphysik, Heidelberg, Germany,
$^6$ Dept. of Physics and Astronomy, Iowa State University, Iowa,
$^7$ SISSA/ISAS, Trieste, Italy,
$^8$ Astroparticle Group, PCC - College de France, Paris, France,
$^9$ CASS/UCSD, La Jolla, California,
$^{10}$ Astronomical Obs. of Torino, Pino Torinese, Italy,
$^{11}$ Penn State University, Pennsylvania,
$^{12}$ Centre d'Etudes Nucleaires de Bordeaux-Gradignan, France,
$^{13}$ Astronomical Obs., Univ. of Perugia, Perugia, Italy,
$^{14}$ Dept. of Physics, Univ. of Insubria, Como, Italy,
$^{15}$ Space Telescope Science Institute, Baltimore, Maryland
}

\begin{document}

\maketitle

\begin{abstract}
We present the preliminary results of a long BeppoSAX observation of the
BL Lac object Mkn501 carried out in June 1999.  The source was fainter
than found during the BeppoSAX pointings of 1997 and 1998, but is still
detected with a good signal-to-noise ratio up to $\sim$40 keV. The X-ray
spectrum in the energy range 0.1-40 keV, produced through synchrotron
radiation, is steeper than in the previous years, it is clearly curved,
and peaks (in $\nu F_{\nu}$) at $\sim$0.5 keV.  This energy is much lower
than
those at which the synchrotron component was found to peak in 1997 and
1998.  Some intraday variability suggests that activity of the source on
small time scales accompanies the large long time scale changes of
brightness and spectrum.  
\end{abstract}

\section{Scientific goal of the program}

The radio-to-$\gamma$-ray spectra of blazars ($\nu F_\nu$) are typically
``double-humped", with the first peak commonly attributed to synchrotron
radiation within a relativistic jet and the second to inverse Compton
scattering of relativistic electrons off synchrotron or ambient soft photons
(Ulrich, Maraschi \& Urry 1997, and references therein). In X-ray
bright BL Lacs (XBL) the
synchrotron maximum generally occurs in or close to the X-ray band, and
the inverse Compton emission peaks above the GeV spectral region extending
in some cases to the TeV band, as observed so far for several sources by
ground based Cherenkov telescopes.  Among these is the nearby ($z = 0.034$)
BL Lac object Mkn~501, one of the brightest blazars at UV, X- and
$\gamma$-ray energies, and a typical XBL according to the numerous
multiwavelength data taken prior to the BeppoSAX launch.  Repeated
observations with the satellites ASCA and XTE have detected large amplitude
X-ray variability at different time scales, often correlated with strong
activity in the TeV band (e.g., Kataoka et al. 1999; Sambruna et al.
2000). 

BeppoSAX observations of Mkn~501 in April 1997, during an outburst,
revealed a completely new behavior.  In fact, the joint LECS, MECS and PDS
spectra showed that at that epoch the synchrotron component peaked at 100
keV or higher energies (Fig. 1).  Correspondingly the source was extremely bright
in the TeV band
and exhibited rapid flares (Catanese et al. 1997; Aharonian et
al. 1999). The flux at 10 keV was an order of
magnitude higher than the historical level, while around 1 keV the flux was
only moderately brighter than usual. The X-ray spectrum hardens with increasing
intensity, and the peak energy of the synchrotron component varies by more
than two decades with respect to the quiescent state, a unique behavior in blazars, if compared with the
variations of the peak energy exhibited by similar sources, never exceeding
an order of magnitude with respect to quiescence (Fossati et al. 2000;
Giommi et al. 1999). A discussion of the SED of Mkn 501 in the frame of
the SSC model is reported in Tavecchio \& Maraschi (this volume).

Further BeppoSAX observations in April-May 1998 showed that the synchrotron
peak energy was located at $\sim$20 keV (Fig. 1), indicating
a
decrease of an order of magnitude with respect to the previous year
(Pian et al. 1999). The simultaneously measured TeV flux was also much lower than in 1997 
(Krawczinsky, priv comm.) 
However, the fact that, despite the radiation losses, the synchrotron peak
was still at such high energies one year after the huge outburst, clearly
indicates the presence of very powerful, efficient and continuously active
mechanisms of particle energization and acceleration in this source. 

\begin{figure}
\hskip -1.1cm
\vskip -1.9cm
\psfig{file=pian_hr.ps, width=5.7cm, height=7.5cm}
\vskip -7.6cm
\hskip 5.5cm
\psfig{file=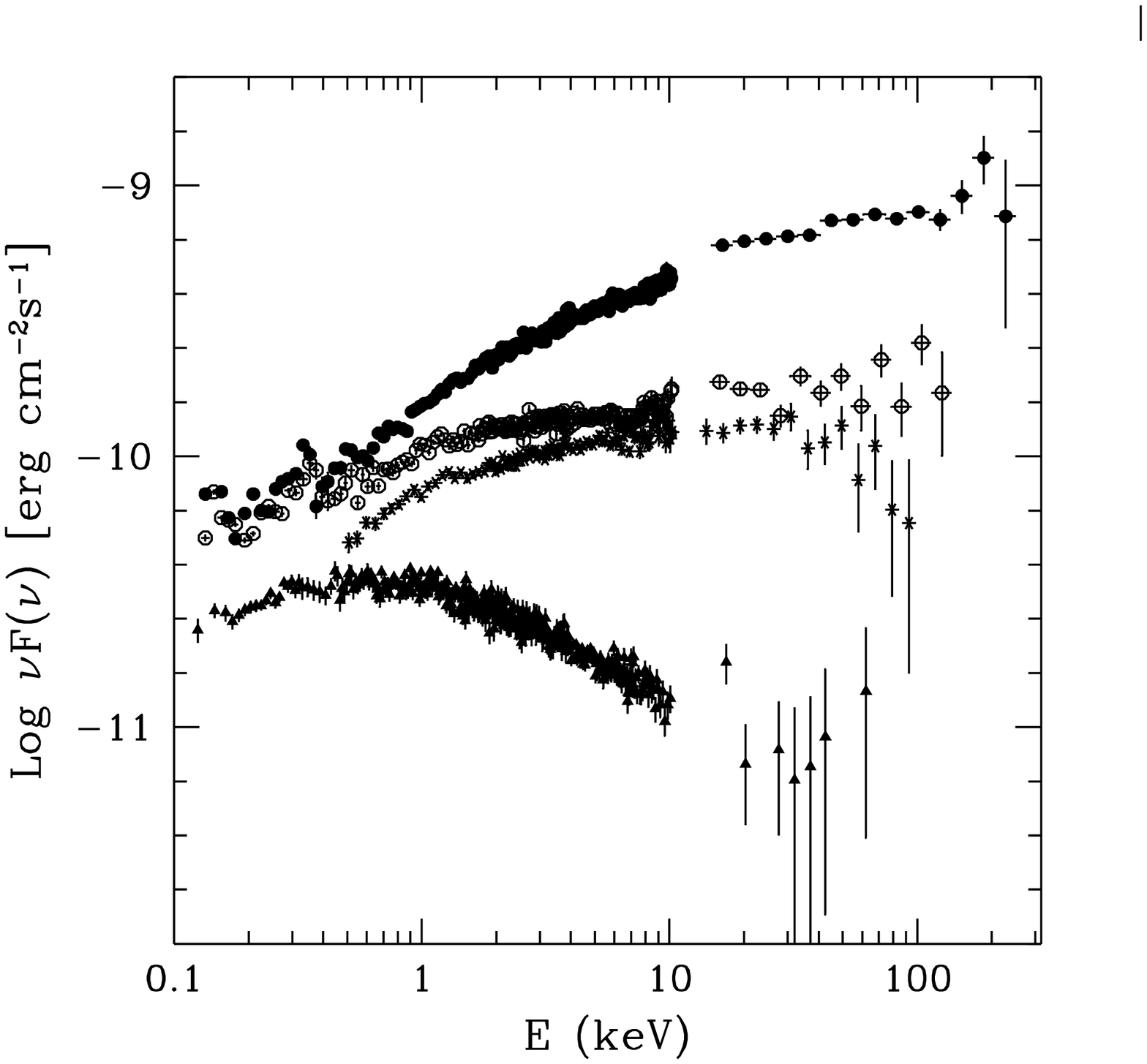, width=7.2cm, height=8.cm}
\caption[]{Left panel: Light curves (0.1-2 keV and 4-10 keV) and Hardness
Ratio of Mkn 501 during the 1999 June observation. Right panel: History of the 
X-ray spectrum of Mkn 501. From top to bottom: 1997 Apr. 16 (filled circles), 1997 Apr. 7 
(open circles), 1998 Apr. 28 (stars) and 1999 June (triangles). }
\vskip -.2cm
\end{figure}

\section{BeppoSAX Observations: June 1999 Campaign}

We have carried out observations with BeppoSAX in June 1999, 
simultaneously with TeV
Cherenkov telescopes and coordinated with XTE, 
to investigate the

\begin{itemize}

\item
X-ray variability of the continuum on short (hours and sub-hour) time
scales;

\item
the presence of possible time lags between soft and hard X-ray
emission;

\item
the correlation between X-ray and gamma-ray flux and
spectral index variations. 

\end{itemize}

A single, 180 ks long BeppoSAX pointing has been performed between 1999
June 10, 23:01:18 UT and June 16, 02:11:08 UT, during one of 
the longest observing windows of the Cherenkov telescopes Whipple, HEGRA, 
CAT and CELESTE. The X-ray data have been cleaned
and linearized at the BeppoSAX Science Data Center (SDC). Spectra and
light curves have been extracted from the images with
the standard XSELECT package.  For the spectral analysis, we used the background
files and response matrices distributed by the SDC.

\section{Results}

The source has been clearly detected by the BeppoSAX LECS and MECS
instruments, and by the PDS up to 40 keV. The flux exhibits
variability of up to 20-30\% in amplitude on time scales of 10-12 hours or
more (Fig. 1). 

The June 1999 flux level at 1 keV is similar to the one observed for Mkn
501 prior to 1997 (``historical" state), namely more than a factor of 2
fainter than detected by BeppoSAX in April-May 1998 and in 1997 April 7,
and an order of magnitude fainter than observed by BeppoSAX during the
outburst of 1997 April 16 (Fig. 1).

The 0.1-40 keV spectrum is steeper than found in 1997 and 1998 and is
progressively steepening with energy. It is not well fitted ($\chi ^2 _r\sim
5$) by a single power-law plus Galactic absorption ($N_H=1.73 \times
10^{20}$ cm$^{-2}$), therefore two power-laws have been used to fit the
data.  The fitted energy break is around ~1 keV and the spectral indices
are $\alpha _1=0.89\pm 0.03$ and $\alpha _2=1.44\pm 0.02$ (errors are at
90\% confidence level; $\chi ^2 _r=1.4$). Therefore,
the X-ray spectrum is consistent with being produced with a unique
emission component, which we identify with synchrotron radiation.

In a $\nu f_\nu$ representation, the X-ray spectrum appears to peak at the
energy of $\sim$1 keV, which can be identified with the maximum of the
synchrotron component.  The comparison with the BeppoSAX spectra of the
previous years (see Fig. 1) indicates that in 1999 the synchrotron peak
has shifted toward lower energies, following the trend
already noted in 1997 and 1998, when the peak was observed at $>$100 keV
and
$\sim$20 keV, respectively. 

Optical and TeV coverage simultaneous with the present BeppoSAX campaign
was limited due to bad weather.  The optical flux level is similar to that
usually observed for Mkn 501.  At TeV energies, only marginal
detections have been obtained on each night.  The analysis of these data
is still underway. 

Our preliminary conclusions are that 

\begin{itemize}

\item
The energy at which the synchrotron component peaks can vary on long term
by a large amount (more than a factor of $\sim 200$), in correspondence
with long time scale large amplitude flux variations (2 orders of magnitude at
10 keV).

\item 
This ``shift" in energy takes place on much longer time scales (years)
than the synchrotron cooling times at X-ray energies, estimated from
multiwavelength energy distribution fitting (e.g. Tavecchio \& Maraschi,
this volume).  This further indicates that electrons are continuously
accelerated, while the high state is gradually turning to quiescence.

\item
The low TeV flux in June 1999 indicates that the TeV
flux variations are probably correlated with the X-ray variations.

\end{itemize}

\begin{acknowledgements}
We thank the BeppoSAX SDC and Mission Planning Team for their support of
this project.
\end{acknowledgements}

\end{document}